\documentclass[aps,prl,amsmath,amssymb,twocolumn,superscriptaddress,letterpaper]{revtex4}
\usepackage{dcolumn}
\usepackage{graphicx}
\usepackage{epstopdf}
\usepackage{verbatim}
\usepackage{amssymb}
\usepackage{amsmath}
\usepackage{subfigure,wrapfig}
\usepackage{color}

\begin{document}

\title{Metamaterial Broadband Angular Selectivity}

\author{Yichen Shen,$^{1\dagger}$ Dexin Ye,$^{2}$ Zhiyu Wang,$^{2}$ Li Wang,$^{3}$ Ivan Celanovic, $^{1}$ \\ Lixin Ran,$^{2}$ John D Joannopoulos,$^{1}$ and Marin Solja\v{c}i\'{c}$^{1}$}
\affiliation{
\normalsize{$^{1}$Research Laboratory of Electronics, Massachusetts Institute of Technology,}\\
\normalsize{Cambridge, MA 02139, USA}\\
\normalsize{$^{2}$Laboratory of Applied Research on Electromagnetics (ARE), Zhejiang University,} \\ 
\normalsize{Hangzhou, 310027, China}\\
\normalsize{$^{3}$Key Laboratory of Artificial Micro- and Nano- structures of Ministry of Education, Wuhan University,}\\
\normalsize{Wuhan 430072, China}\\
\normalsize{$^\dagger$To whom correspondence should be addressed; E-mail:  ycshen@mit.edu.}
}

\begin{abstract}

We demonstrate how broadband angular selectivity can be achieved with stacks of one-dimensionally periodic photonic crystals, each consisting of alternating isotropic layers and effective anisotropic layers, where each effective anisotropic layer is constructed from a multilayered metamaterial. We show that by simply changing the structure of the metamaterials, the selective angle can be tuned to a broad range of angles; and, by increasing the number of stacks, the angular transmission window can be made as narrow as desired. As a proof of principle, we realize the idea experimentally in the microwave regime. The angular selectivity and tunability we report here can have various applications such as in directional control of electromagnetic emitters and detectors.

\end{abstract}

\maketitle
Light selection based purely on the direction of propagation has long been a scientific challenge \cite{PhysRevLett.106.123902,Atwater_angular2013,PhysRevB.85.205430}. Narrow band angularly selective materials can be achieved by metamaterials \cite{Schwartz:03} or photonic crystals \cite{Upping2010102}; however, optimally, an angularly selective material-system should work over a broadband spectrum. Such a system could play a crucial role in many applications, such as directional control of electromagnetic wave emitters and detectors, high efficiency solar energy conversion \cite{doi:10.1117/12.921773, Bermel:10}, privacy protection \cite{privacyfilter:2003}, and high signal-to-noise-ratio detectors. 

Recent work by Shen \textit{et al} \cite{Shen28032014} has shown that one can utilize the characteristic Brewster modes to achieve broadband angular selectivity. The key concept in that work was to tailor the overlap of the bandgaps of multiple one-dimensional isotropic photonic crystals, each with a different periodicity, such that the bandgaps cover the entire visible spectrum, while visible light propagating at the Brewster angle of the material-system \textit{does not experience any reflections}. Unfortunately, for an isotropic-isotropic bilayer system, the Brewster angle is determined solely by the two dielectric constants of these materials; hence, it is \textit{fixed} once the materials are given. Furthermore, among naturally occurring materials, one does not have much flexibility in choosing materials that have the precisely needed dielectric constants, and therefore the available range of the Brewster angles is limited. For example, the Brewster angle at the interface of two dielectric media (in the lower index isotropic material) is always larger than 45$^{\circ}$. In many of the applications mentioned above, it is crucial for the material-system to have an arbitrary selective angle, instead of only angles larger than 45$^{\circ}$. Furthermore, the ability to control light would be even better if the selective angle could be tuned easily and dynamically.

In this report,  we build upon earlier work by Hamam \textit{et al.} \cite{PhysRevA.83.035806}, who pointed out that an angular photonic band gap can exist within anisotropic material-systems, to introduce a design that can in principle achieve a broadband angular selective behavior at \textit{arbitrary} incident angles. Furthermore, it can be easily fabricated in the microwave regime, and even possibly in the optical regime. As a proof of principle, an experiment in the microwave regime for the normal-incidence-selective case is reported.

\begin{figure}[htbp]
\begin{center}
\includegraphics[width=1.2in]{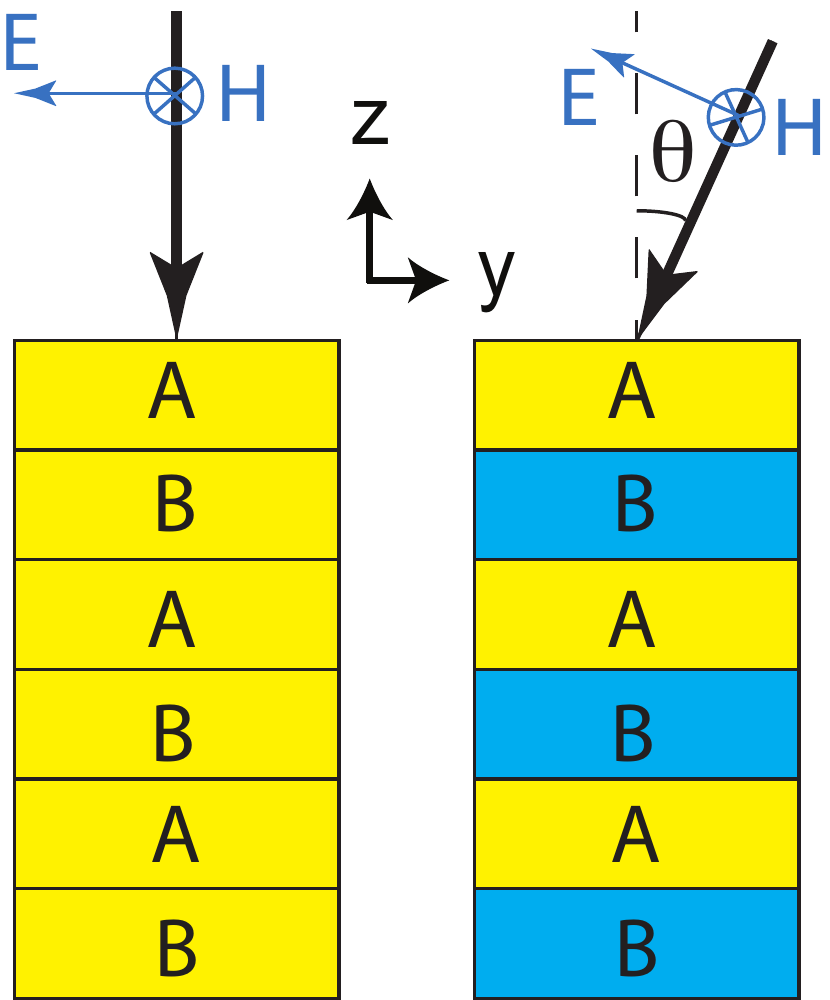}
\caption{Effective index for $p$-polarized light in an isotropic(A) - anisotropic(B) multilayer system. Left panel: all the layers have the same index. Right panel: the change in incident angle leads to a change in the index of the anisotropic (B) material layers.}
\label{fig:figure1}
\end{center}
\end{figure}
\vspace{-18pt}

Our angular selective material-system is built starting from a one-dimensionally periodic photonic crystal (1D-PhC), as shown in figure 1; it consists of isotropic layers (A) and anisotropic layers (B). The key idea rests on the fact that $p$-polarized light is transmited without any reflection at an ``effective'' Brewster angle of the isotropic-anisotropic interface.

To show this quantitatively, the reflectivity of $p$-polarized light with a propagating angle $\theta_{i}$ (defined in isotropic material) at an isotropic-anisotropic interface can be computed using a transfer matrix method \cite{azzam1977}:

\begin{equation}
R_p=\left|\frac{n_xn_z\cos{\theta_{i}}-n_{iso}(n_z^2-n_{iso}^2\sin{\theta_{i}}^2)^{\frac{1}{2}}}{n_xn_z\cos{\theta_{i}}+n_{iso}(n_z^2-n_{iso}^2\sin{\theta_{i}}^2)^{\frac{1}{2}}}\right|^2
\label{eqn:iso-ani}
\end{equation}
where $n_x=n_y$ and $n_z$ are the refractive indices of the anisotropic material at the ordinary and extraordinary axes, respectively, and $n_{iso}$ is the refractive index of the isotropic material.

Therefore, the Brewster angle, $\theta_i=\theta_B$, can be calculated by setting $R_p=0$, giving:
\begin{equation}
\tan{\theta_B}=\sqrt{(\frac{\epsilon_z}{\epsilon_{iso}})\cdot\left[\frac{\frac{\epsilon_x}{\epsilon_{iso}}-1}{\frac{\epsilon_z}{\epsilon_{iso}}-1}\right]}
\label{eqn:brewster}
\end{equation}

\begin{figure}[htbp]
\begin{center}
\includegraphics[width=3.2in]{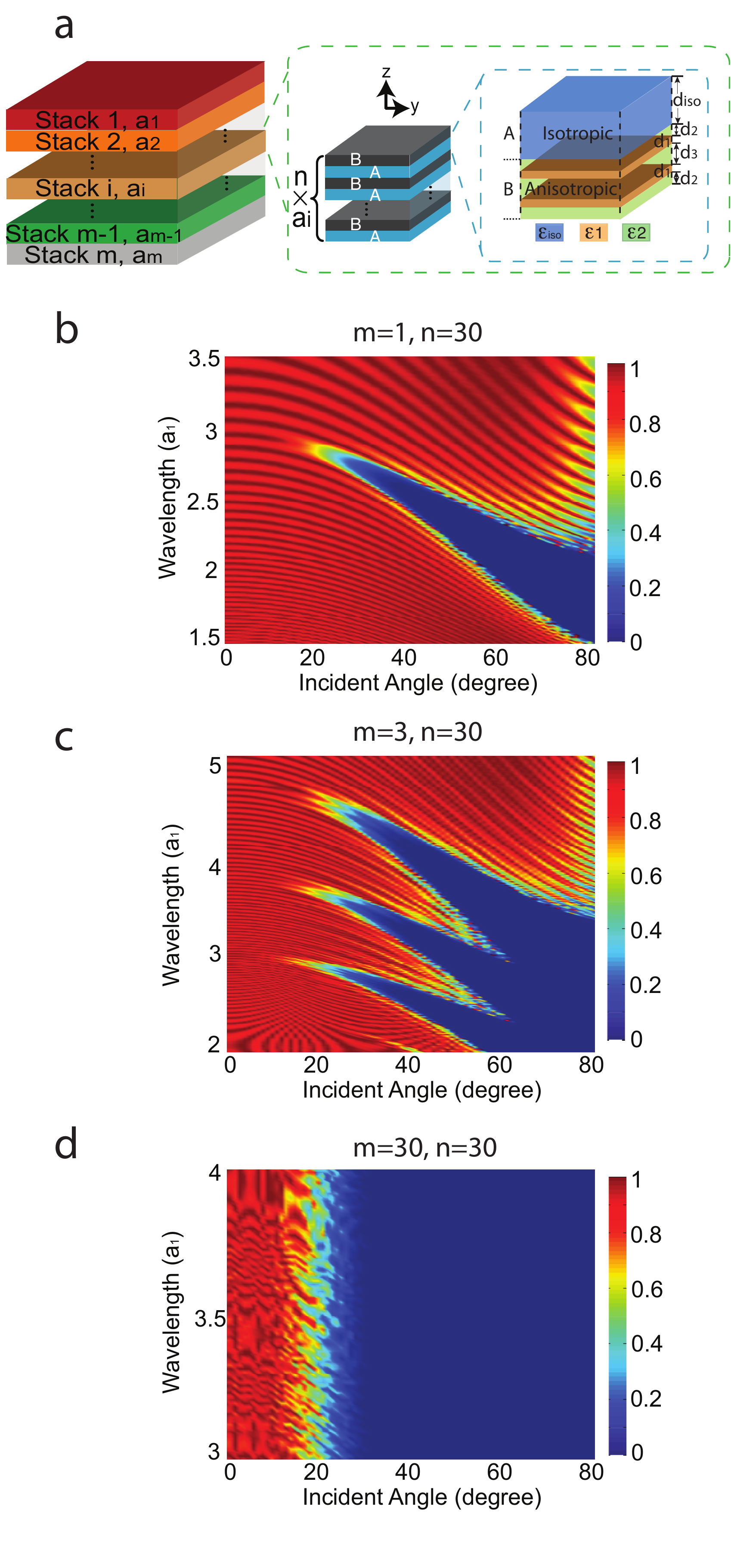}
\vspace{-18pt}
\caption{\textbf{Metamaterial Behavior.} a, Schematic illustration of a stack of isotropic-anisotropic photonic crystals. Layer A is an isotropic medium, layer B is an effective anisotropic medium consists of two different isotropic media with dielectric constants $\epsilon_1$ and $\epsilon_2$. b, $p$-polarized transmission spectrum for a 30 bilayer structure with $\{\epsilon_{iso}, \epsilon_1, \epsilon_2\}=\{2.25, 10, 1\}$, and $r=6.5$. The unit of $y$ axis - $a_1$ is the periodicity of the structure. c, $p$-polarized transmission spectrum of 3 stacks of 30 bilayer structures described in part b. The periodicities of these stacks form a geometric series $a_i = a_1q^{i-1}$ with q = 1.26, where $a_i$ is the periodicity of i$^{th}$ stack. d, $p$-polarized transmission spectrum of 30 stacks of 30 bilayer structures described in part b. The periodicities of these stacks form a geometric series $a_i = a_1q^{i-1}$ with q = 1.0234, where $a_i$ is the periodicity of i$^{th}$ stack.}
\label{fig:figure2}
\end{center}
\end{figure}

To demonstrate how the angular photonic bandgap can be opened with an isotropic-anisotropic photonic crystal, we begin by considering an example that achieves broadband angular selectivity at normal incidence. From Eqn.~\ref{eqn:brewster}, in order to have $\theta_B=0$, we need to choose the permittivity of the isotropic material to be equal to the $xy$ plane-component permittivity of the anisotropic material, that is
\begin{equation}
\epsilon_{isotropic}=\epsilon_x=\epsilon_y
\label{eqn:normal_requirement}
\end{equation}
In an anisotropic material, the analytical expressions for the effective refractive index are given by \cite{boyd2003}:

\begin{equation}
\frac{1}{n_{e}(\theta)^{2}}=\frac{\sin^{2}\theta}{\widetilde{n_{e}}^{2}}+\frac{\cos^{2}\theta}{\widetilde{n_{o}}^{2}},
\label{eqn:4}
\end{equation}
and
\begin{equation}
n_{o}(\theta)=\widetilde{n_{o}},
\label{eqn:2}
\end{equation}
where $\theta$ is the angle between the $\hat{z}$ axis (Fig.~2(a)) and $\overrightarrow{k}$ in the material. $n_{e}(\theta)$ is the effective refractive index experienced by extraordinary waves, $n_{o}(\theta)$ is the effective refractive index experienced by ordinary waves, $\widetilde{n_{e}}^{2}=\frac{\epsilon_{z}^{B}}{\epsilon_{0}}$ is the refractive index experienced by the $z$ component of the electric field, and $\widetilde{n_{o}}^{2}=\frac{\epsilon_{x}^{B}}{\epsilon_{0}}=\frac{\epsilon_{y}^{B}}{\epsilon_{0}}$ is the refractive index experienced by the $x$ and $y$ components of the electric field.

At normal incidence, for $p$-polarized light, the effective dielectric constant of the anisotropic layers $n_e(0)^2=\widetilde{n_{o}}^{2}=\epsilon_x$ matches the isotropic layer $\epsilon_{iso}$; therefore no photonic band gap exists, and all the normal incident light gets transmitted ($R_p=0$ in Eqn.~\ref{eqn:iso-ani}). On the other hand, when the incident light is no longer normal to the surface, the $p$-polarized light has $E_{z}\neq0$, and experiences an index contrast $n_{A}^{p}(\theta)=\sqrt{\epsilon_{iso}}\neq n_{B}^{p}=n_e(\theta)$ (Fig.~1). As a result, a photonic band gap will open. Furthermore, we notice that as $\theta$ gets larger, the $\widetilde{n_{e}}$ term in Eqn.~\ref{eqn:4} becomes more important, hence the size of the bandgap increases as the propagation angle deviates from the normal direction. The band gap causes reflection of the $p$-polarized incident light, while the $s$-polarized light still has $E_{z}=0$, so it remains as an ordinary wave experiencing no index contrast $n_{A}^{s}=n_{B}^{s}$; hence, $s$-polarized light will be transmited at all angles.

The method described above provides an idealistic way of creating an angular photonic bandgap. However, in practise it is hard to find a low-loss anisotropic material, as well as an isotropic material whose dielectric constants exactly match that of the anisotropic material. In our design, we use a metamaterial to replace the anisotropic layers in Fig.~1, as shown in Fig.~2(a). Each metamaterial layer consists of several high-index ($\epsilon_1=10$) and low-index ($\epsilon_2=\epsilon_{air}=1$) material layers. We assume that each layer has a homogeneous and isotropic permittivity and permeability. When the high-index layers are sufficiently thin compared with the wavelength, the effective medium theorem allows us to treat the whole system as a single anisotropic medium with the \textit{effective} dielectric permittivity tensor $\{\epsilon_x,\epsilon_y,\epsilon_z\}$ \cite{Bergman1978377}: 

\begin{equation}
\epsilon_x=\epsilon_y=\frac{\epsilon_1+r\epsilon_2}{1+r}
\label{eqn:1}
\end{equation}

\begin{equation}
\frac{1}{\epsilon_z}=\frac{1}{1+r}\left(\frac{1}{\epsilon_1}+\frac{r}{\epsilon_2}\right)
\label{eqn:2}
\end{equation}
where $r$ is the ratio of the thickness of the two materials $\epsilon_1$ and $\epsilon_2$: $r=\frac{d_2}{d_1}$.

For example, in order to achieve the normal incidence angular selectivity, we need the dielectric permittivity tensor of the anisotropic material to satisfy Eqn.~\ref{eqn:normal_requirement}. For the isotropic material (A) layers, we need to choose $\epsilon_{iso}$ that lies between $\epsilon_1$ and $\epsilon_2$. For definiteness, we choose a practical value of $\epsilon_{iso}=2.25$ (common polymers). From Eqn.~\ref{eqn:1} and Eqn.~\ref{eqn:2}, with material properties $\epsilon_1=10$ and $\epsilon_2=1$, and the constraint $\epsilon_x=\epsilon_y=\epsilon_{iso}=2.25$, we can solve for $r$, obtaining $r=6.5$.

Using the parameters calculated above and with a 30-bilayer structure, the transmission spectrum of $p$-polarized light at various incident angles is calculated using the Transfer Matrix Method \cite{hecht2008optics}, and the result is plotted in Fig.~2(b). In Fig.~2(b) the wavelength regime plotted is much larger than $d_1$; in such a regime, the light interacts with layer B as if it is a homogeneous medium, and experiences an effective anisotropic dielectric permittivity.

\begin{figure}[htbp]
\begin{center}
\includegraphics[width=3.5in]{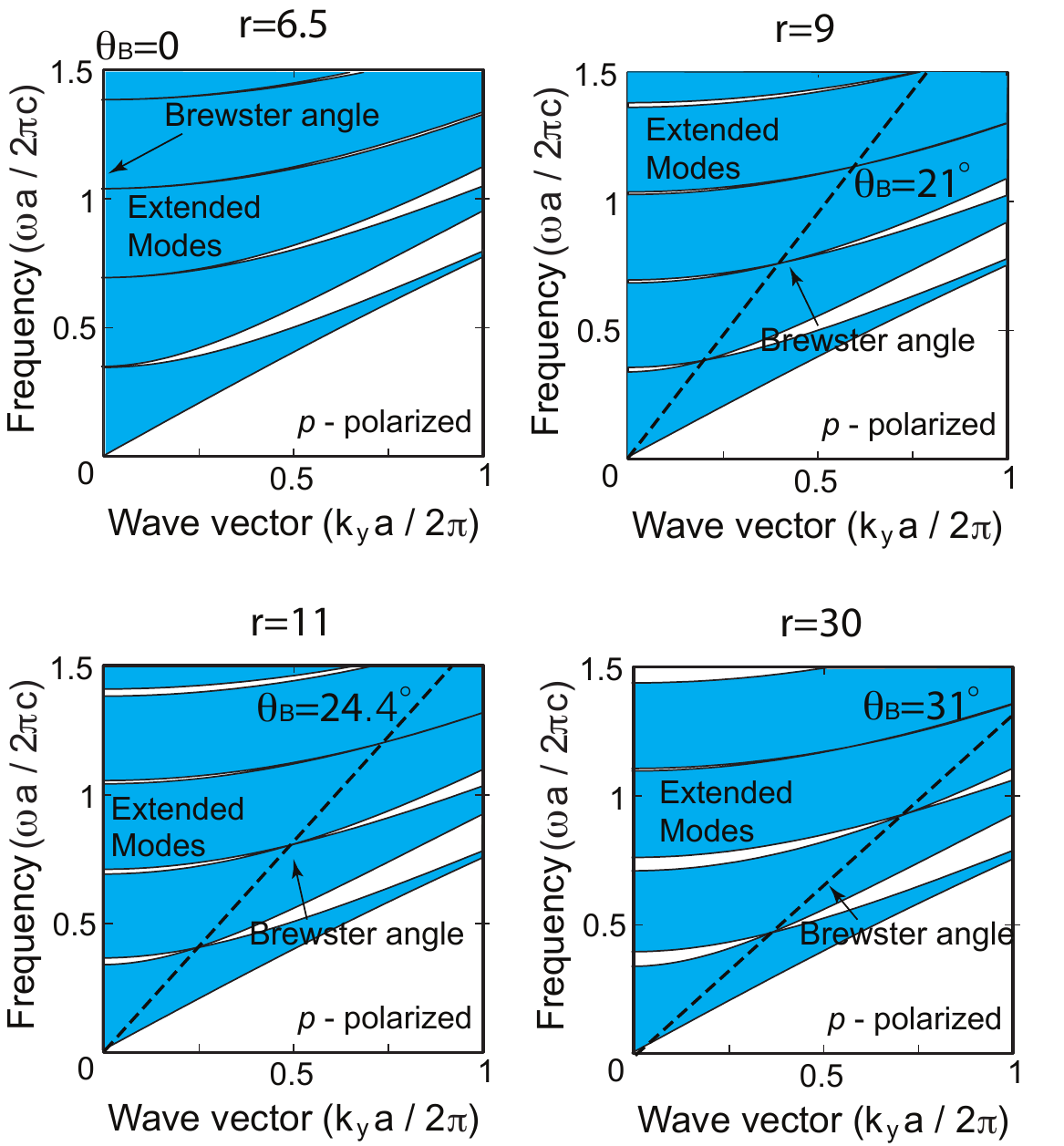}
\vspace{-18pt}
\caption{\textbf{Photonic band diagram of one-dimensional isotropic-anisotropic quarter-wave stacks.} Modes that are allowed to propagate (so-called extended modes) exist in the shaded regions; no modes are allowed to propagate in the white regions (known as bandgaps). The Brewster angles are marked with a dash line in each case.}
\label{fig:figure3}
\end{center}
\end{figure}

One can enhance the bandwidth of the angular photonic bandgap by stacking more bilayers with different periodicities \cite{Xin2002heterostructure, Xu2010stacking}. In Fig.~2(c,d), we present the stacking effect on the transmission spectrum for $p$-polarized light. When we have a sufficient number of stacks, a broadband angular selectivity (bandwidth $>$ 30\%) at normal incidence can be achieved. 

\begin{figure}[htbp]
\begin{center}
\includegraphics[width=3.5in]{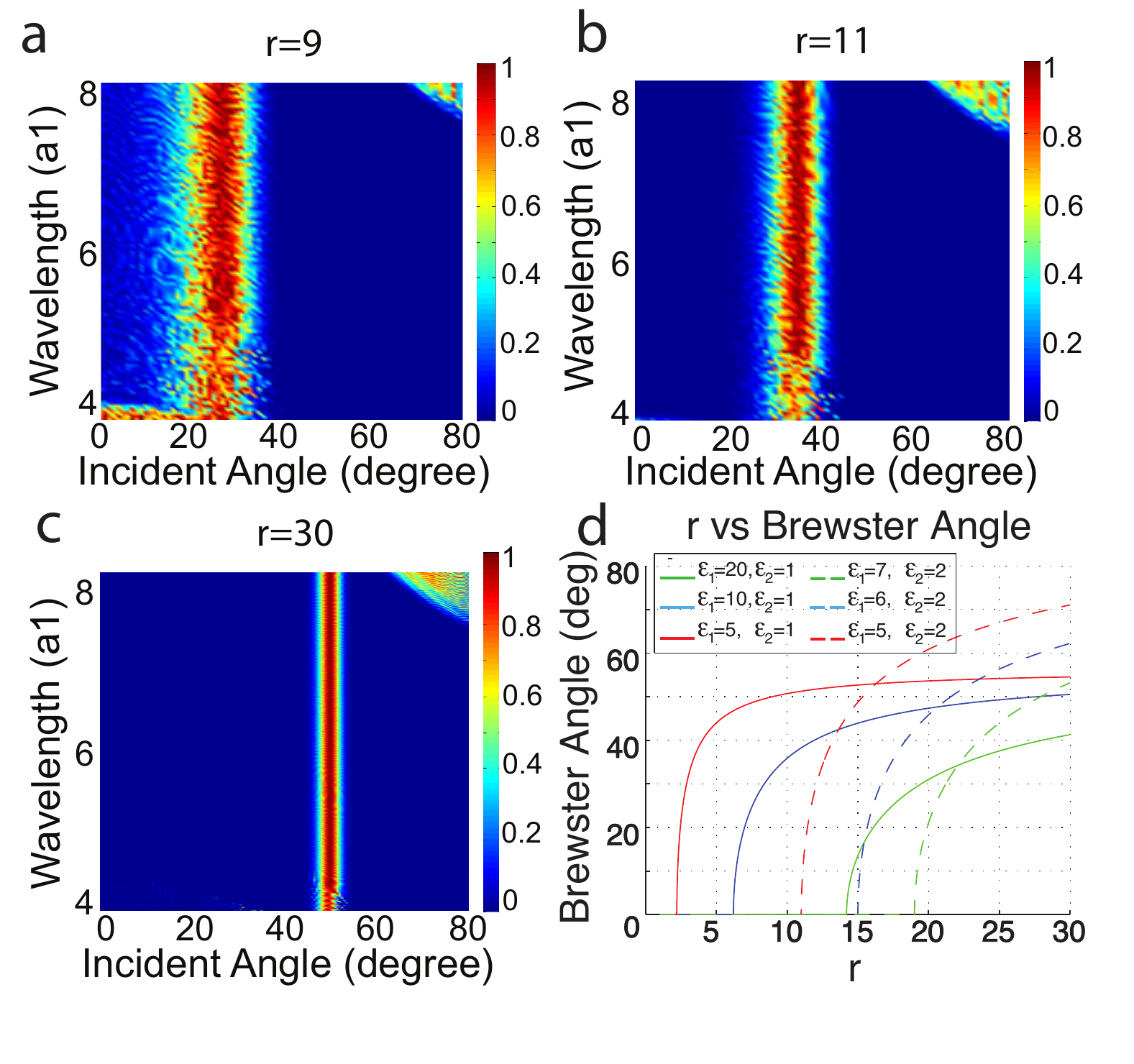}
\vspace{-18pt}
\caption{\textbf{Angular selectivity at oblique angles.} Same materials and structure as illustrated in Fig.~2, n=m=30, and $q=1.0373$ but with different $r$. In a,b,c we used $\epsilon_1=10$, $\epsilon_2=1$, $\epsilon_{iso}=2.25$. a, $r=9$, and $\theta_B=24^{\circ}$. b, $r=11$, and $\theta_B=38^{\circ}$. c, $r=30$,  and $\theta_B=50^{\circ}$. d, Dependence of the Brewster angle on $r$ for various values of $\epsilon_1$ and $\epsilon_2$. Solid and dashed lines correspond respectively to $\epsilon_2$ for air and $\epsilon_2$ for PDMS.}
\label{fig:figure4}
\end{center}
\end{figure}

In general, the Brewster angle for isotropic-anisotropic photonic crystals depends on $\epsilon_x$, $\epsilon_z$ and $\epsilon_{iso}$ (Eqn.~\ref{eqn:brewster}). In our metamaterial-system, it depends strongly on $r$. Substituting Eqn.~\ref{eqn:1} and Eqn.~\ref{eqn:2} into Eqn.~\ref{eqn:brewster}, we get:
\begin{equation}
\theta_B(r)=\arctan\left[\sqrt{\frac{\epsilon_1'\epsilon_2'(\epsilon_1'+r\epsilon_2'-1-r)}{(1+r)\epsilon_1'\epsilon_2'-\epsilon_2'-\epsilon_1'r}}\right ]
\label{eqn:brewster_final}
\end{equation}
where $\epsilon_1'=\frac{\epsilon_1}{\epsilon_{iso}}$, and $\epsilon_2'=\frac{\epsilon_2}{\epsilon_{iso}}$. From Eqn.~\ref{eqn:brewster}, we can see that in order to have a non-trivial Brewster angle, we need $\epsilon_{iso}$ to be larger than $\max\{\epsilon_x,\epsilon_z\}$ or smaller than $\min\{\epsilon_x,\epsilon_z\}$; otherwise there will be no Brewster angle.

The result in Eqn.~\ref{eqn:brewster_final} shows that it is possible to adjust the Brewster angle by changing the ratio $r=\frac{d_1}{d_2}$, or, by changing the spacing distance $d_2$ when everything else is fixed. In Fig.~3, we show the photonic band diagrams \cite{joannopoulos2011photonic} of a simple anisotropic-isotropic quarter-wave stack. The band diagrams are calculated with preconditioned conjugate-gradient minimization of the block Rayleigh quotient in a planewave basis, using a freely available software package \cite{Johnson2001:mpb}. The dielectric tensor of the anisotropic material in each band diagram is calculated by Eqn.~\ref{eqn:1} with $r=6.5$, $r=9$, $r=11$ and $r=30$, respectively. It is clear that the Brewster angle increases as we increase $r$: when $r\rightarrow\infty$, the Brewster angle (defined in the isotropic layer) approaches $\theta_B=\arctan\sqrt{\frac{\epsilon_2}{\epsilon_{iso}}}$ \cite{hecht2008optics}. Note that if $\epsilon_2$ is some soft elastic material (such as PDMS or air), one can simply vary $r$ easily by changing the distance $d_2$ in real time, and hence varying the Brewster angle accordingly. Such \textit{tunability} of the Brewster angle does not exist in conventional (non-metamaterial) isotropic-isotropic or isotropic-anisotropic photonic crystals, where the Brewster angle depends solely on the materials' dielectric properties.

Similar to what we obtained in Fig.~2, we can enhance the bandwidth of angular selectivity by adding stacks with different periodicities. The transmission spectra of metamaterial-systems with $m=n=30$ (see Fig.~2a) and different $r$'s were calculated with Transfer Matrix Method \cite{hecht2008optics} and plotted in Fig.~4(a,b,c) respectively. Note that due to the inherent properties of Eqn.~1, the angular selective window gets narrower as the Brewster angle increases. Furthermore, the dependence of the Brewster angle on $r$ is presented in Fig.~4(d). As r gets larger, we see a rapid increase in the Brewster angle, which eventually plateaus, approaching the isotropic-isotropic limit, $\theta_B=\arctan\sqrt{\frac{\epsilon_2}{\epsilon_{iso}}}$.

\begin{figure}[htbp]
\begin{center}
\includegraphics[width=3.2in]{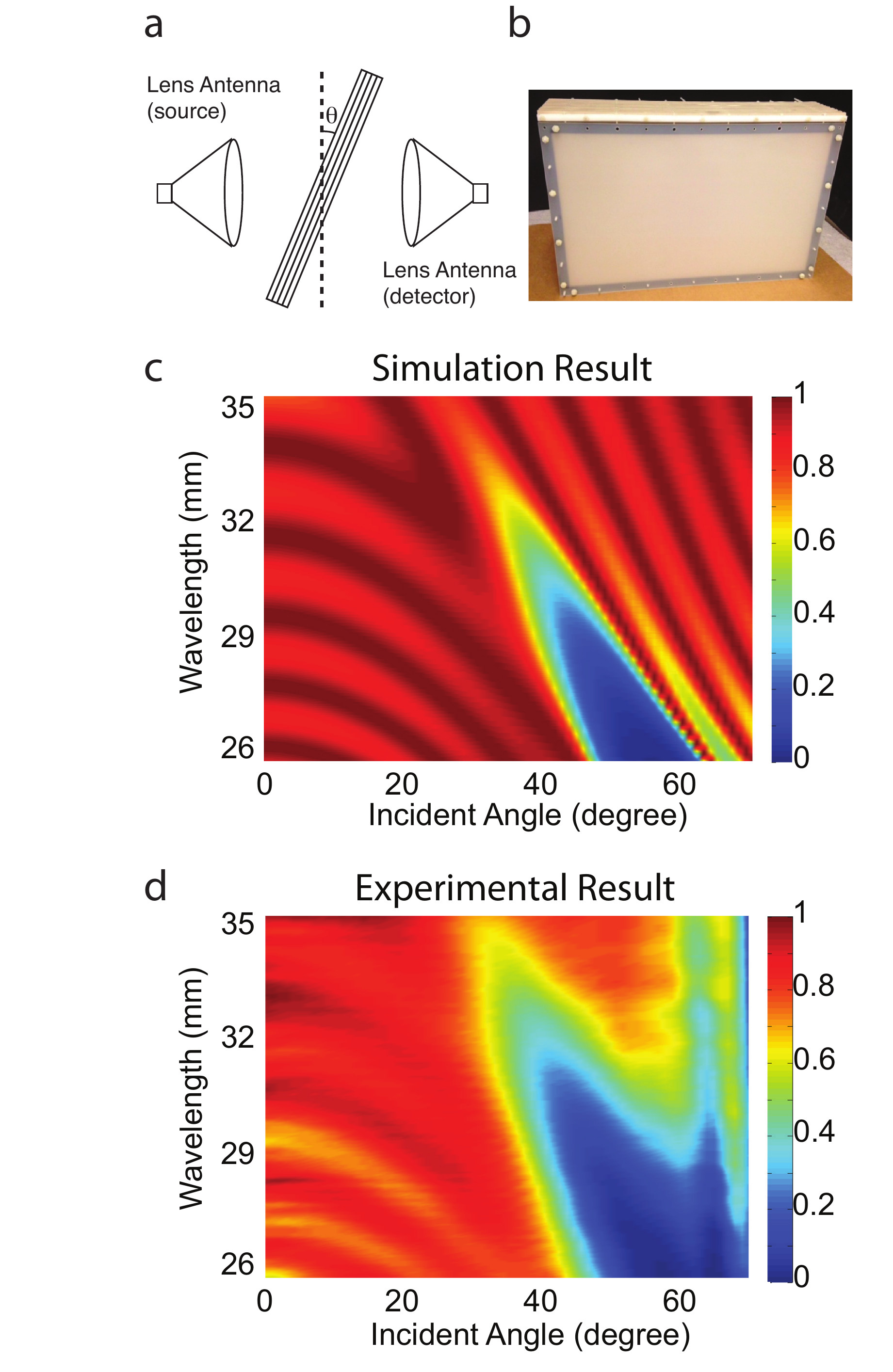}
\vspace{-12pt}
\caption{\textbf{Experimental verification} a, A schematic illustration of the experimental setup. b, Photo of the fabricated sample. c, d, Comparison between $p$-polarized transmission spectrum of Transfer Matrix Method, and the experimental measurements.}
\label{fig:figure5}
\end{center}
\end{figure}

To show the feasibility of the above-mentioned method, we present an experimental realization in the microwave regime. Since our goal here is only to demonstrate the concept, we kept the experimental setup simple. We implemented the geometry design in Fig.~2a, using Rogers R3010 material ($\epsilon_1=10$), air ($\epsilon_2=1$), and polypropylene ($\epsilon_{iso}=2.25$, Interstate Plastics). The thickness of each layer is chosen to be $\{d_{iso},d_1,d_2,d_3\}=\{3.9,0.5,1.6,3.9\}mm$. A simple 12-period structure ($m=12$, $n=1$ case in Fig.~2a) was made (Fig.~5b). With the experimental setup shown in Fig.~5a, the transmission spectrum for $p$-polarized light was measured in the wavelength range from $26mm$ to $35mm$. For incident angles less than 60 degrees, the experimental result (Fig.~5d) agrees well with the simulation (analytical) result calculated from Transfer Matrix Method \cite{hecht2008optics} (Fig.~5c). For larger incident angles, the finite-sized microwave beam spot picks up the edge of the sample, which causes the transmission to deviate from the theoretical simulation; by using bigger samples, one should be able to resolve this issue.

In the present work, we have introduced a basic concept of using stacks of isotropic-anisotropic one-dimensionally periodic photonic crystals to achieve broadband angular selectivity of light. The key idea in our design is using the generalized Brewster angle in hetero-structured photonic crystals. Due to the limited choice of natural anisotropic materials, the method of using metamaterials to create an effective anisotropic medium is proposed. With the proposed material-system, we demonstrate the possibility of tuning the selective angle in a wide range, simply by modifying the design of the material-system with a fixed material composition. As a proof of principle, a simple experimental demonstration in the microwave regime was reported.

Compared with previous work in \cite{Shen28032014}, the wide-range angular tunability is one of the core advantages of the metamaterial-system proposed in this report. This feature could enable new applications (in addition to conventional angular selective devices) in the microwave and optical frequency regimes, such as radar tracking and laser scanning \cite{angular:2014}. Furthermore, the mechanism proposed in this paper enables transmission for both polarizations at normal incidence. A natural next step would be to fabricate material-systems with more layers and extend the working frequency regime to the infrared or visible spectra. For example, one can sputter a material-system consisting of SiO$_2$ ($\epsilon_1=2$), PMMA ($\epsilon_{iso}=2.25$) and Ta$_2$O$_5$ ($\epsilon_2=4.33$) to realize the angular selective filter at arbitrary angle in the optical regime. Another potentially interesting feature would be to explore the dynamical tuning of the selective angle.

We thank Dr. Ling Lu for the valuable discussion. This work was partially supported by the Army Research Office through the ISN under Contract No.~W911NF-13-D0001, and Chinese National Science Foundation (CNSF) under Contract No.~61131002. The fabrication part of the effort, as well as (partially) M.S. were supported by the MIT S3TEC Energy Research Frontier Center of the Department of Energy under Grant No. DE-SC0001299.

\bibliography{Yichen_bib}
\bibliographystyle{Science}

\end{document}